\documentclass[%
 aip,
 jap,
 amsmath,amssymb,
 reprint,%
]{revtex4-1}

\usepackage{graphicx}
\usepackage{dcolumn}
\usepackage{bm}

\usepackage[utf8]{inputenc}
\usepackage[T1]{fontenc}
\usepackage{graphicx}
\usepackage{mathptmx}
\usepackage{etoolbox}
\usepackage{siunitx}
\usepackage{physics}
\usepackage[capitalize]{cleveref}
\usepackage[version=4]{mhchem}
\usepackage{lineno}
\usepackage{xcolor, soul}
\usepackage[colorinlistoftodos]{todonotes}

\newcommand{\Rb}{\textsuperscript{87}Rb }

\makeatother
\begin{document}

\preprint{AIP/123-QED}

\title{A Compact, High-Data-Rate, High-Sensitivity GMOT-Based Light-Pulse Atom Interferometer toward Time-Multiplexed, Multi-Axis Inertial Sensing}

\author{Benjamin Smith}
\affiliation{Sandia National Laboratories, Albuquerque, New Mexico 87185, USA}

\author{Adrian Orozco}
\affiliation{Sandia National Laboratories, Albuquerque, New Mexico 87185, USA}


\author{Jongmin Lee}
\email{jlee7@sandia.gov}
\affiliation{Sandia National Laboratories, Albuquerque, New Mexico 87185, USA}
\affiliation{Department of Physics and Astronomy, University of New Mexico,  Albuquerque, NM 87106, USA}

\date{\today}

\begin{abstract}
Light-pulse atom interferometry has emerged as a promising approach for developing quantum inertial and gravity sensors for real-world applications, but achieving robust operation in dynamic environments without sacrificing sensitivity remains challenging. To mitigate performance degradations resulting from vibration-induced noise and mismatches in the spatial overlap between free-falling atoms and light pulses, we demonstrate a first-of-a-kind high-data-rate, high-sensitivity grating magneto-optical trap (GMOT)-based light-pulse atom interferometer (LPAI). This system uses a compact, rugged GMOT architecture, based on an in-vacuum microfabricated grating chip and a single laser-cooling beam, to enable high-data-rate GMOT-based cold-atom generation with sub-Doppler cooling. Our GMOT-based LPAI operates at high data rates (up to \SI{100}{Hz}), while simultaneously achieving both high acceleration sensitivity (\SI{550}{\nano g / \sqrt{Hz}} at \SI{20}{Hz}, \SI{1.49}{\micro g / \sqrt{Hz}} at \SI{45}{Hz}, and \SI{6.31}{\micro g / \sqrt{Hz}} at \SI{70}{Hz}) and good bias stability (\SI{71}{ng}). This compact, rugged architecture paves the way for field-deployable quantum inertial and gravity sensors in high dynamics and for time-multiplexed, multi-axis inertial sensing.
\end{abstract}

\maketitle

\section*{Introduction}
\label{sec:intro}

Light-pulse atom interferometers (LPAIs) \cite{Kasevich91, Bongs19, Geiger20} have been developed for inertial sensing \cite{Gustavson00} and precision gravity measurements. \cite{Peters01} LPAIs have achieved exquisite sensitivity \cite{Dickerson13, Sugarbaker13} and long-term bias stability compared to conventional inertial sensors. Colder atom clouds, which provide longer de Broglie matterwave coherence lengths, and extended interrogation times remain key to improved performance. In stationary laboratories, LPAIs typically rely on complex ultra-cold-atom systems with data rates of 0.01–\SI{0.05}{Hz} and long-baseline vacuum chambers to maximize their interrogation time.\cite{Dickerson13, Sugarbaker13} To reduce size, weight, and power (SWaP) for field deployment, portable cold-atom LPAIs \cite{Geiger11, Barrett16, Bidel18, Menoret18, Wu19, Becker18} operate at data rates of 1–\SI{5}{Hz} and have demonstrated robust performance on aircraft and ships under mild vibration ($\lesssim$ 1-to-3 $\rm{grms}$).

Inertial measurement units (IMUs) are essential for guidance, navigation, and control. They typically contain inertial sensors: a three-axis accelerometer that measures linear acceleration and a three-axis gyroscope that measures angular velocity. Quantum IMUs have therefore been explored for field-deployable applications because of their potential for high sensitivity and improved long-term bias stability. However, mission-relevant conditions (e.g. high dynamics, broadband vibrations, acoustic noise, thermal cycling, and radiation) impose immense demands on the underlying atom interferometers and their subsystems. Low-data-rate LPAI-based quantum IMUs (1–\SI{5}{Hz}) maintain high sensitivity in stationary or gently vibrating platforms but suffer severe degradation under vehicle-induced vibrations \cite{Menoret18} and high-dynamic conditions. \cite{Soh20} Their low data rates, limited measurement bandwidth, and long free-fall times make them susceptible to vibration-induced phase noise, aliasing, and imperfect spatial overlap between the free-falling atoms and the interrogation light pulses. By contrast, conventional IMUs, with data rates of approximately 500–\SI{1500}{Hz}, can operate robustly in harsh dynamic environments but generally lack the long-term bias stability offered by quantum IMUs.

\begin{figure}[b!]
\includegraphics[width=1\columnwidth]{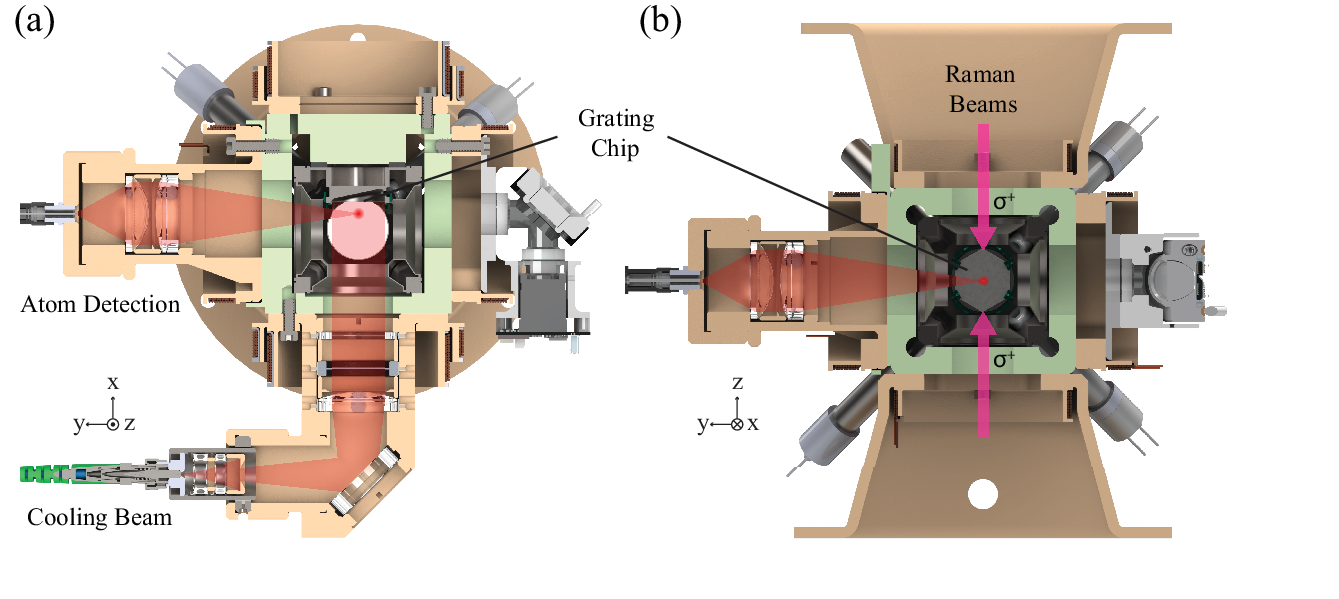}
\caption{Rendered 3D design of the compact grating magneto-optical trap (GMOT) architecture for high-data-rate cold-atom generation with sub-Doppler cooling and a high-data-rate light-pulse atom interferometer (LPAI). (a) Horizontal sectional view of the GMOT physics package, which incorporates an in-vacuum microfabricated diffraction grating and generates sub-Doppler-cooled atoms at data rates up to \SI{100}{Hz}. (b) Vertical sectional view of the GMOT physics package, which shows the microfabricated grating chip, whose surface faces the -x-direction, mounted in a custom titanium vacuum chamber on a 3D-printed structure with a swan-neck flexure. The GMOT-based LPAI employs Doppler-sensitive Raman beams along the z-axis to impart state-dependent photon recoils.}
\label{fig:Sensor_Head}
\end{figure}

To leverage the complementary strengths of both technologies, several groups have developed hybrid inertial sensors \cite{Templier22, Cheiney18} that combine low-data-rate quantum sensors (1–\SI{5}{Hz}) with high-data-rate conventional sensors (500–\SI{1500}{Hz}), rejecting common-mode noise and improving sensitivity and stability under quasi-static or mild-vibration conditions. Nevertheless, the quantum sensor's low data rate still constraints performance in high-vibration, high-dynamics maneuvers. These limitations underscore the critical need for compact, rugged physics packages, \cite{Lee22, Seo26} compact photonic integrated circuit (PIC) laser subsystems, \cite{Kodigala24} support for sustained high-data-rate LPAI operation, and advanced quantum-classical sensor-fusion algorithms that exploit both high data rates and the intrinsic long-term stability of quantum sensors.

In our previous work, \cite{Lee22} we developed a compact, rugged grating magneto-optical trap (GMOT) physics package (\cref{fig:Sensor_Head}) to support field-deployable GMOT-based LPAI operation. The ultra-high vacuum chamber, which forms the core of the physics package, consists of a custom Grade-2 titanium chamber with five anti-reflection-coated fused-silica viewports, non-evaporable getters, and alkali-metal dispensers housed in laser-welded Ti tubing. Titanium was chosen over steel because it is nonmagnetic, has a lower thermal expansion coefficient, and exhibits a much lower hydrogen outgassing rate. The package also includes a copper exhaust port for evacuation, an ion pump for maintaining vacuum, and an in-vacuum microfabricated grating chip mounted with a sintered, 3D-printed titanium-alloy retainer with four swan-neck flexures for a single-beam GMOT operation. 

\begin{figure}[t!]
\includegraphics[width=1\columnwidth]{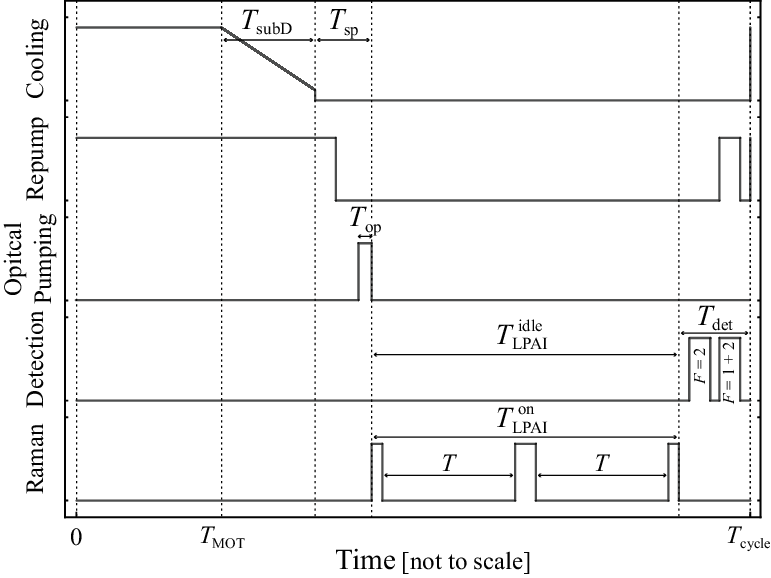}
\caption{Experimental timing diagrams for two experiments: (1) high-data-rate GMOT-based cold-atom generation with sub-Doppler cooling, featuring an idle free-evolution (drop) time $T_\text{LPAI}^\text{idle}$ and no Raman beams; and (2) high-data-rate, high-sensitivity GMOT-based LPAI demonstration, featuring a light-pulse sequence time $T_\text{LPAI}^\text{on}$ (corresponding to a maximum interrogation time of $\approx 2 T$) with Raman beams. The data rate is defined by $1/T_\text{cycle}$. In experiment (1), $T_\text{cycle}^{(1)} = T_\text{MOT} + T_\text{subD} + T_\text{LPAI}^\text{idle} + T_\text{ovhd}$. In experiment (2), $T_\text{cycle}^{(2)} = T_\text{MOT} + T_\text{subD} + T_\text{LPAI}^\text{on} + T_\text{ovhd}$. Here $T_\text{MOT}$ is the GMOT reload and recapture time, $T_\text{subD}$ is the sub-Doppler cooling time, $T_\text{sp}$ is the state preparation time, $T_\text{det}$ is the atom detection time, and $T_\text{ovhd}$ is the overhead time ($T_\text{sp}$ plus $T_\text{det}$). $S_{F=2}$ is the signals from the $F=2$ state, and $S_{F=1+2}$ are  and from the total atom population.}
\label{fig:Timing}
\end{figure}

Unlike a conventional six-beam MOT, these GMOT-based LPAI architectures \cite{Lee22, Seo26} has a single incoming cooling beam incident on a microfabricated, in-vacuum reflective grating chip. \cite{Vangeleyn10, Nshii13, Lee13, Imhof17, McGilligan17, Barker19, Sitaram20, McGehee21, Seo21} The grating splits the beam into three first-order beams, forming a tetrahedral, sub-Doppler magneto-optical trap (MOT) operating at data rates up to \SI{20}{Hz}. We housed the grating in a custom titanium vacuum chamber \cite{Little21} equipped with fused-silica windows, alkali dispensers, and a \SI{2}{L/S} ion pump. In a parallel variant using sapphire windows and passive getter pumps, \cite{Little21} we demonstrated continuous cold-atom loading for over five years without active pumping. Finally, by adding counter-propagating, cross-linearly-polarized Doppler-sensitive Raman beams, we realized a GMOT-based LPAI at \SI{10}{Hz} with a statistical uncertainty of \SI{2}{\micro g} as determined by time-scanned fringe-chirp fitting. \cite{Lee22} 

As shown in \cref{fig:Sensor_Head}, this single-beam approach greatly simplifies both light delivery and vacuum-chamber configurations; the reduced number of cooling beams and improved optical access shrink the physics-package footprint by \SI{2}{L} and enable a streamlined opto-mechanical structure for low-phase-noise Raman interrogation pulses, substantially improving robustness under vibration and high-dynamic environments.

We also developed a novel PIC-based laser system architecture featuring a \SI{1560}{nm}-to-\SI{780}{nm} frequency-doubling approach. Using a single master laser and silicon-photonic single-sideband modulators, we individually demonstrated each phase of a single-cycle experiment for LPAI operation: laser cooling and trapping, Doppler-sensitive Raman interrogation, and state-selective detection. \cite{Kodigala24} A four-channel silicon-photonic single-sideband modulator chip (also developed at Sandia) will eventually integrate all of these functions into a single PIC. Together, these advances will pave the way toward a heterogeneously integrated, PIC-based laser system for real-world quantum inertial and gravity sensing.

\begin{figure}[h]
\includegraphics[width=1\columnwidth]{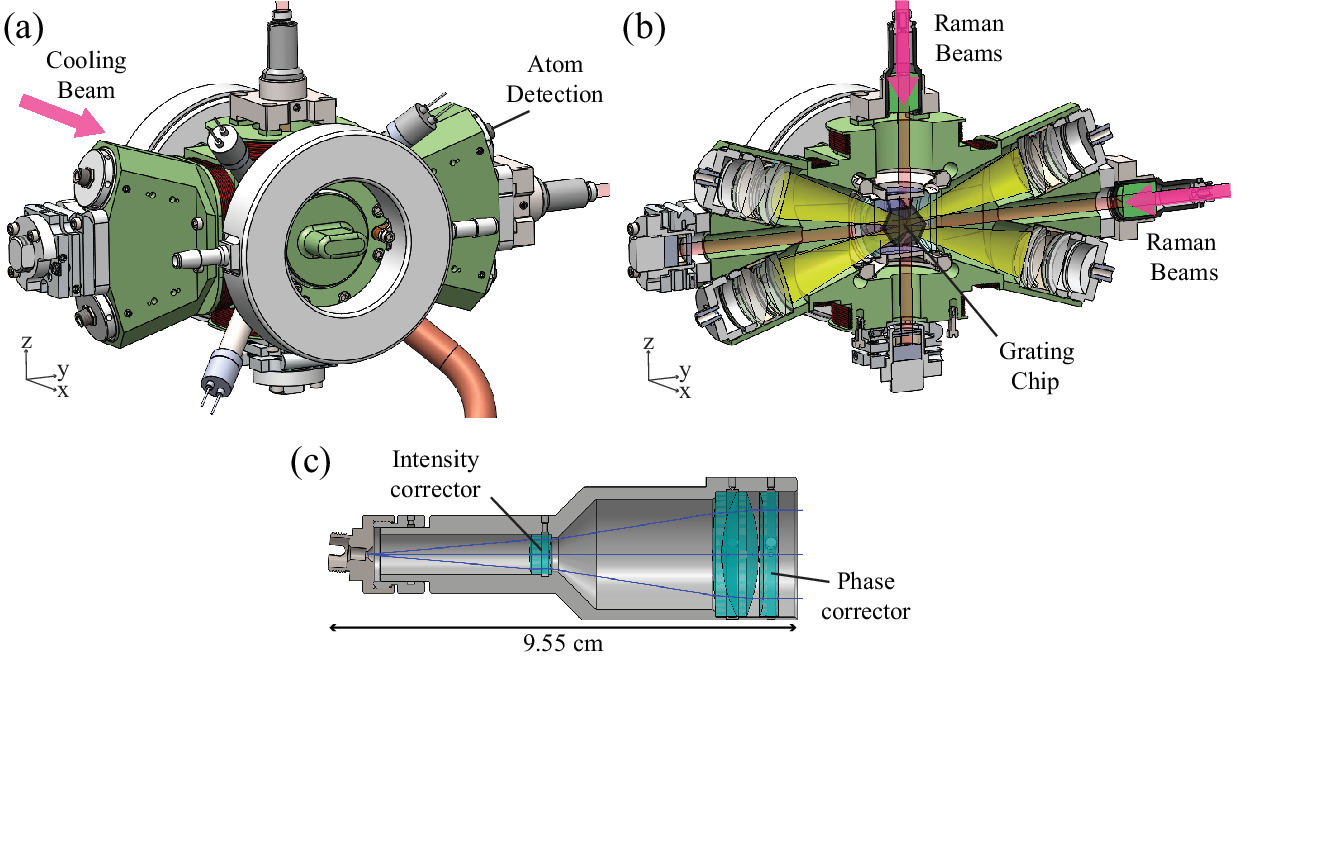}
\caption{Rendered 3D design of the compact GMOT architecture for time-multiplexed, multi-axis acceleration sensing. (a) The GMOT physics package is equipped with two-axis Raman beams along the y- and z-axes for two-axis acceleration sensing. The GMOT cooling beam propagates along the x axis. Four detection channels are arranged with offsets along both the z- and x-axes. (b) Vertical sectional view of the GMOT physics package, with the surface of the in-vacuum microfabricated grating chip facing the -x-direction. The view illustrates how four partially overlapping atom-detection channels expand the overall detection volume and enable high-dynamic operation of a GMOT-based LPAI. (c) Custom 10-cm-long flat-top beam shaper for expanding and collimating the GMOT cooling beam using custom intensity- and phase-correction optics.}
\label{fig:Multi-Axis_Sensor}
\end{figure}

In our current setup, we use a PIC-compatible laser system based on a previously demonstrated \SI{1560}{nm}-to-\SI{780}{nm} frequency-doubling approach. \cite{Lee22} For these experiments (\cref{fig:Timing}), we have also added a standalone 795-nm optical-pumping laser to increase the atomic population in the desired initial state and a tapered amplifier to boost the GMOT cooling-beam power, thereby increasing the GMOT atom number and improving recapture efficiency. These modifications support high-data-rate (up to \SI{100}{Hz}) GMOT-based cold-atom generation with sub-Doppler cooling, for which efficient light delivery is critical: a large-diameter Gaussian beam truncated by the optical window can improve spatial uniformity at the grating chip but delivers cooling power inefficiently, while the GMOT capture volume (\SI{\approx 0.17}{cm^3}) is smaller than that of a conventional MOT. To enhance the light-delivery efficiency at the in-vacuum microfabricated grating chip to more than \SI{70}{\%}, we developed and tested a custom 10-cm-long flat-top beam shaper that expands and collimates the beam for future prototypes.

In this paper, we investigate the potential of a compact, high-data-rate, high-sensitivity GMOT-based LPAI toward time-multiplexed, multi-axis inertial sensing. First, we characterized the recapture efficiency (the fraction of cold atoms reused each cycle) by measuring its dependence on the idle free-evolution (drop) time and data rate, and demonstrated operation up to \SI{100}{Hz} in our current laboratory setup. Second, by performing Allan deviation analysis at various data rates, we determined the acceleration sensitivity and bias stability. Third, for each data rate, we varied the interrogation time $T$ and identified an optimal $T$ that maximizes sensitivity. Finally, we show that the overall optimum arises from a trade-off between the statistical gain of higher data rates and the $T^2$-scaling of atom interferometric sensitivity.

This study indicates that, for a GMOT-based LPAI to achieve high sensitivity at data rates above \SI{70}{Hz}, the capture volume during Doppler cooling may need to be increased to improve the overall recapture efficiency. Inspired by collimated- or diverging-beam MOTs, which exhibit larger capture volumes than GMOTs and therefore provide better recapture efficiency for cold-atom reuse, additional helper beams could be introduced along the horizontal and vertical sensing axes (\cref{fig:Multi-Axis_Sensor}) only during Doppler cooling (GMOT reload and recapture). These beams would expand the effective capture volume to \SI{\gtrsim 1}{cm^3}, supporting reliable high-data-rate GMOT-based cold-atom generation with sub-Doppler cooling toward time-multiplexed, multi-axis acceleration sensing.

This design demonstrates that high-data-rate cold-atom interferometry offers a compelling solution to meet the platform-robustness requirements under real-world conditions, enabling precision inertial navigation with reduced GPS reliance. In principle, high-data-rate cold-atom interferometry \cite{Butts11, McGuinness12, Rakholia14} minimizes the deleterious effects of relative motion between freely released atoms in vacuum and the interrogation light pulses delivered by the physics package. To achieve practical demonstration, we made two key breakthroughs: (1) generating sub-Doppler-cooled atomic clouds at high data rates using a GMOT and proposing the helper beams discussed above as a potential means to preserve high atom-recapture efficiency in future implementations, and (2) implementing a high-data-rate, high-sensitivity GMOT-based LPAI, making it less vulnerable to real-world perturbations while still maintaining performance despite the usual $T^2$ sensitivity scaling. 

Our short-$T$ GMOT-based LPAI, operating at high data rates, offers a statistical advantage in sensitivity and minimizes spatial-overlap mismatches between free-falling atoms and light pulses, thereby increasing the dynamic range. Its short-baseline geometry also reduces the overall physics-package size. Moreover, by sharing high-data-rate GMOT-based cold-atom clouds among multiple axes (or across disparate sensors), we reduce the total number of laser systems needed for laser cooling and trapping, and atom-light interactions, naturally enabling time-multiplexed, multi-axis sensing in a compact form factor.

\cref{tab:timing_sequence} and \cref{tab:GMOT_LPAI_seq} detail our sequence:
\begin{table}[h]
\caption{Typical timing settings for two experiments, with the timing diagram shown in \cref{fig:Timing}, for (1) high-data-rate GMOT-based cold-atom generation with sub-Doppler cooling and (2) high-data-rate, high-sensitivity GMOT-based LPAI demonstration.}
    \centering
    \begin{tabular}{l l l c}
        \hline
        Seq. & Description & Duration \\ 
        \hline
        $T_\text{MOT}$  & GMOT reload and recapture &  3.6 to \SI{500}{ms}  \\
        \hline
        $T_\text{subD}$ & Sub-Doppler cooling  &  2 to \SI{2.5}{ms} \\
        \hline
        $T_\text{LPAI}^\text{idle}$ & Idle free evolution (drop)  &  1 to \SI{34}{ms} \\
        \hline
        $T_\text{LPAI}^\text{on}$       & Light-pulse sequence (max. interrog.)  &  1 to \SI{34}{ms} \\
        \hline
        $T_\text{ovhd}$  & Overhead ($T_\text{sp} + T_\text{det}$)                             &  \SI{1.85}{ms}\\
        \hline
        $T_\text{cycle}^{(1)}$ & $T_\text{MOT} + T_\text{subD} + T_\text{LPAI}^\text{idle} + T_\text{ovhd}$ & \text{A single cycle of (1)}\\
        \hline
        $T_\text{cycle}^{(2)}$ & $T_\text{MOT} + T_\text{subD} + T_\text{LPAI}^\text{on} + T_\text{ovhd}$ & \text{A single cycle of (2)}\\
        \hline
    \end{tabular}
    \label{tab:timing_sequence}
\end{table}

Hybrid inertial sensors with complementary sensor-fusion schemes could combine the strength of quantum IMUs, high sensitivity and long-term bias stability, with those of conventional IMUs, high data rates and large dynamic ranges. Such systems could leverage quantum-derived stability across a broader operating envelop of an inertial navigation system. However, one remaining challenge is that the real-world dynamics and vibrations can degrade LPAI performance (especially in low-data-rate, long-interrogation-time systems) underscoring the need to enhance the robustness of both physics packages and laser subsystems and to advance sensor-fusion algorithms. Furthermore, incorporating map-matching techniques (e.g. magnetic- or gravity-aided navigation) can mitigate IMU errors and drift, further reduce reliance on GPS, and extend the duration and range of fully GPS-free navigation trajectories.

In this paper, we experimentally assess the potential of a compact, high-data-rate, high-sensitivity GMOT-based LPAI as a building block for future time-multiplexed, multi-axis inertial sensing. By optimizing sub-Doppler-cooling, atom recapture, and interrogation time at data rates up to \SI{100}{Hz}, we achieve acceleration sensitivities of \SI{550}{\nano g / \sqrt{Hz}} at \SI{20}{Hz}, \SI{1.49}{\micro g / \sqrt{Hz}} at \SI{45}{Hz}, and \SI{6.31}{\micro g / \sqrt{Hz}} at \SI{70}{Hz}, along with a bias stability of \SI{71}{ng}. These results represent a significant step toward compact, rugged quantum inertial and gravity sensors suitable for real-world deployment.

\section*{Results}

We first demonstrate high-data-rate cold atom generation using a GMOT. A single laser beam incident on a microfabricated grating inside the vacuum chamber creates a tetrahedral MOT, and cycle-to-cycle recapture readies sub-Doppler-cooled atoms in a few milliseconds---enabling measurement rates up to \SI{100}{Hz}, well above conventional MOT systems (1--\SI{5}{Hz}). Once the atoms are cooled, a precisely timed light-pulse sequence splits, redirects, and recombines atomic wavepackets to produce matterwave interference fringes whose phase shifts directly encode acceleration or angular velocity. Running this interferometry cycle at 20--\SI{100}{Hz} boosts sampling and achieves better sensitivity despite the short interrogation time (\SI{\leq 17}{ms}). By sweeping the interrogation time at different data rates we confirmed the expected trade-off: at \SI{20}{Hz} (interrogation times from \SI{1}{ms} to \SI{17}{ms}), the optimal sensitivity of \SI{550}{\nano g / \sqrt{Hz}} occurs at \SI{12}{ms}---where high data rates improve statistical sensitivity, whereas longer interrogation times enhance per-shot sensitivity, yielding an optimal operating point. In addition, short interrogation times at high data rates raise the low-frequency cutoff of the atom interferometer transfer function, enhancing its robustness against vibrations. These results suggest that our compact GMOT architecture---implementing a high-data-rate, high-sensitivity GMOT-based LPAI---could meet the speed and precision requirements for inertial navigation in high-dynamics environments.

\subsection*{High-Data-Rate GMOT-Based Cold-Atom Generation with Sub-Doppler Cooling}

In our GMOT, high-data-rate cold atom generation relies on tightly timed cooling, repump, and detection pulses within a single cycle of duration $T_\text{cycle}^{(1)} = T_\text{MOT} + T_\text{subD} + T_\text{LPAI}^\text{idle} + T_\text{ovhd}$, where $T_\text{MOT}$ is the GMOT reload time; $T_\text{subD}$ is the \SI{2.5}{ms} required for sub-Doppler cooling to \SI{\sim 3}{\micro\K} via a frequency ramp ($-3.33\,\Gamma$ to $-28.4\,\Gamma$) and an intensity ramp from \SI{100}{\%} to \SI{10}{\%};  $T_\text{LPAI}^\text{idle}$ is the idle free-evolution (drop) time (from 1 to \SI{34}{ms}) during which no light pulses are applied, reserved for future light-pulse atom-interferometry sequences; and $T_\text{ovhd}$ covers the overhead time, including state preparation $T_\text{sp}$ (using a $\pi$-polarized $D_1$ optical pumping beam propagating along the +y-axis, to prepare atoms in the $|F = 2, m_F = 0\rangle$ state, quantization axis along the z-axis; see \cref{fig:Sensor_Head}) and normalized atom detection $T_\text{det}$, which is comprised of a first pulse that measures the $F = 2$ population and a second pulse (with the repump beam) that measures the total $F = 1 + 2$ population. The normalized signal grows with longer $T_\text{MOT}$ (more initial atoms) and falls off as $T_\text{LPAI}^\text{idle}$ grows (fewer atoms recaptured). The quantization axis is left on during atom detection.

\begin{figure}[t!]
\includegraphics[width=1\columnwidth]{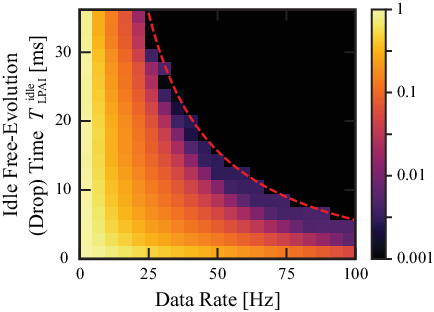}
\caption{Atom recapture efficiency $\beta$ of sub-Doppler-cooled GMOT atoms as a function of data rate and the idle free-evolution (drop) time $T_\text{LPAI}^\text{idle}$. Here, $T_\text{LPAI}^\text{idle}$ excludes the MOT loading time $T_\text{MOT}$, the sub-Doppler cooling time $T_\text{subD}$ (= \SI{2.5}{ms}), and the overhead time $T_\text{ovhd}$ (= $T_\text{sp} + T_\text{det}$ = \SI{1.85}{ms}). The red dashed curve marks the locus $T_\text{MOT}$ = \SI{0}{ms}, i.e. $T_\text{LPAI}^\text{idle} = T_\text{cycle}^{(1)} - (T_\text{subD} + T_\text{ovhd})$}
\label{fig:GMOT_Recapture}
\end{figure}

As shown in \cref{fig:GMOT_Recapture}, we measured the atom recapture efficiency $\beta = N_{eq}/N_{\infty}$, where $N_{eq}$ is the steady-state atom number under cyclic loading at data rate, and $N_{\infty} (= \frac{R_{load}}{\Gamma_{loss}} = 1.8 \times 10^6$) is the asymptotic atom number under continuous loading. The GMOT is characterized by a loading rate $R_{load} = 5.42 \times 10^6$ $\text{atoms/s}$ with a loss rate $\Gamma_{loss}$ of 2.962 $\text{s}^{-1}$, so that $N_{\infty}$ is reached within 2 $\text{s}$. We determine $\beta$ via fluorescence detection for various combinations of data rate and $T_\text{LPAI}^\text{idle}$ (\cref{fig:GMOT_Recapture}) by adjusting $T_\text{MOT}$ while holding $T_\text{subD}$ constant as \SI{2.5}{ms}; $T_\text{ovhd}$, which includes state $T_\text{sp}$ and $T_\text{det}$, is \SI{1.85}{ms}. In each measurement, we subtract background scattered light and optimize the APD signal by adjusting the collection-optic height for the chosen drop time. Only those cycles satisfying $T_\text{LPAI}^\text{idle} \leq T_\text{cycle}^{(1)} - (T_\text{subD} + T_\text{ovhd})$ are physically allowed; the equality corresponds to the red dashed boundary ($T_\text{MOT} =$ \SI{0}{ms}) in \cref{fig:GMOT_Recapture}. At longer idle free-evolution (drop) times, a certain fraction of atoms are further away from the GMOT center and fall out of its capture volume before the reloading process begins, reducing the atom recapture efficiency $\beta$.  Likewise, higher data rates leave less time per cycle for recapture, which also lowers the recapture fraction.

\subsection*{High-Data-Rate, High-Sensitivity GMOT-Based Light-Pulse Atom Interferometer (LPAI)}

Our high-data-rate, high-sensitivity GMOT-based LPAI operation relies on two key elements: recapturing atoms from one cycle to the next, and delivering tightly timed cooling, repump, optical-pumping, detection, and Raman pulses within a single cycle of duration $T_\text{cycle}^{(2)} = T_\text{MOT} + T_\text{subD} + T_\text{LPAI}^\text{on} + T_\text{ovhd}$, where $T_\text{LPAI}^\text{on}$ sets the maximum integration time (e.g. $T_\text{LPAI}^\text{on} \approx 2 T$). 

Recapturing atoms from cycle to cycle \cite{Butts11, McGuinness12, Rakholia14} reduces $T_\text{MOT}$ compared to loading from background vapor and maximizes the available $T_\text{LPAI}^\text{on}$. When data rate $\gg 1$, the short cycle time keeps the reference frame effectively stationary during free fall and also ensures that each light pulse remains well overlapped with the atoms, preserving high-contrast fringes and maintaining sensitivity under high dynamic conditions.

The sensitivity $S$ of an LPAI scales as $S \propto \frac{1}{k_{eff} |g| T^2} \frac{1}{\sqrt{N_{AI}} \cdot \sqrt{\text{data rate}}}$, where $k_{eff}$ is the effective wave number of the stimulated Raman transitions (related to photon recoils), $g$ is the local gravitational acceleration, $N_{AI}$ is the number of atoms in the atom interferometer, and $T$ is the interrogation time. \cite{Adams23} Conventional long-$T$ LPAI operation is premised on $S \propto 1/T^2$, which indicates that a long-$T$ LPAI operating at low data rates yields higher sensitivity. However, because $S \propto 1/\sqrt{\text{data rate}}$, increasing the data rate actually reduces statistical uncertainty and thus partly compensates for shorter $T$. Here, we report a sensitivity of \SI{550}{\nano g/\sqrt{Hz}} (\SI{20}{Hz}, $T = $\SI{12}{ms}), extracted from the Allan deviation ADEV with a \SI{1}{s} integration time (since higher data rate provides more samples within a given integration time, lowering noise even at short $T$).

\begin{figure}[h!]
\includegraphics[width=1\columnwidth]{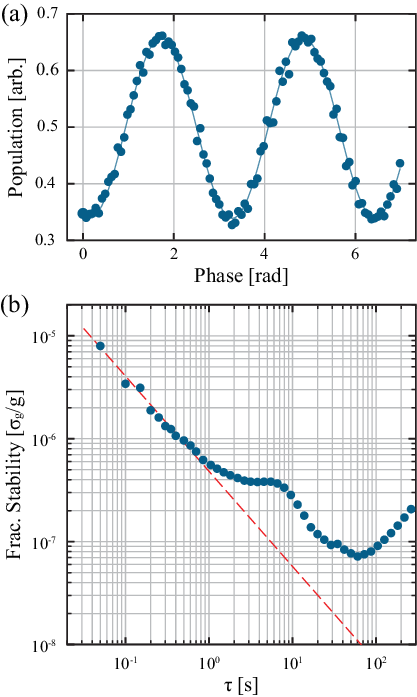}
\caption{Demonstration of a GMOT-based LPAI at \SI{20}{Hz} data rate. (a) Phase-scanned LPAI fringe with a \SI{31.4}{\%} contrast. (b) Allan deviation (ADEV) measurement showing a sensitivity of \SI{550}{\nano g/\sqrt{Hz}} and a bias stability of \SI{71}{\nano g} at \SI{20}{Hz} with $T$ = \SI{12}{ms}.}
\label{fig:GMOT_LPAI_Allan}
\end{figure}

During LPAI operation, counter-propagating, Doppler-sensitive Raman beams (quantization axis along the z-axis; see \cref{fig:Sensor_Head}) drive two-photon stimulated Raman transitions in a three-level $\lambda$ system, coherently coupling internal atomic states and imparting state-dependent photon recoils that generate matterwave interference sensitive to gravitational and inertial forces. The light-pulse sequence $\frac{\pi}{2} \rightarrow T \rightarrow \pi \rightarrow T \rightarrow \frac{\pi}{2}$ splits, redirects, and recombines the atomic wavepackets. Consequently, each $\pi/2$ pulse acts as a beam splitter, and the $\pi$ pulse acts as a mirror for the atomic wavepackets.

Building on this insight, we experimentally recorded an atom-interferometer fringe by scanning the RF phase $\varphi$ of the final $\pi/2$ Raman pulse in the sequence $\frac{\pi}{2} \rightarrow T \rightarrow \pi \rightarrow T \rightarrow \frac{\pi}{2} (\varphi)$ (\cref{fig:GMOT_LPAI_Allan}a). We then held a phase $\varphi$ at the mid-fringe operating point, where the atom interferometer is most sensitive to phase shifts. \cref{fig:GMOT_LPAI_Allan}b shows the ADEV of the mid-fringe atomic population signal, from which we extracted the LPAI's acceleration sensitivity and bias stability. These measurements were performed at the linear operating point of the fringe (where the two ground-state populations are equal), so that small external accelerations produced proportional changes in the detected signal. Because the fringe's signal-to-noise ratio determines the minimum resolvable phase shift, it also defines the achievable acceleration sensitivity.

\begin{figure}[h!]
\includegraphics[width=1\columnwidth]{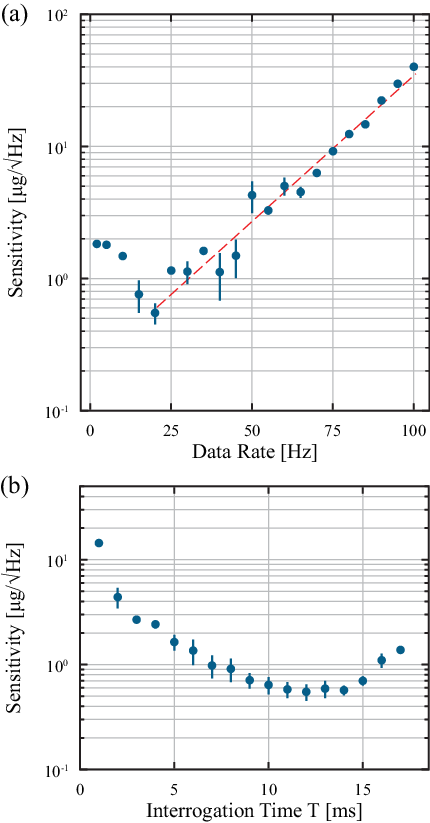}
\caption{Performance metrics of the high-data-rate GMOT atom interferometer. (a) Optimized sensitivity as a function of data rate. The best sensitivity is 550 $\pm$ 10 (statistical) $\pm$ 100 (systematic) [\SI{}{ng}] at \SI{20}{Hz}. The red dashed exponential curve fit in the same plot has the form $s(x) = A \cdot e^{x/b}$, where the fit parameters are A = 220(40) \SI{}{\micro g /\sqrt{Hz}} and b = 19.6(7) Hz. The sensitivities are optimized with $T$ for each data rate individually. (b) Sensitivity is given as a function of $T$ at a \SI{20}{Hz} data rate; the optimal sensitivity occurred at $T =$ \SI{12}{ms}.}
\label{fig:GMOT_LPAI_Perf}
\end{figure}

The statistical uncertainty in the interferometer sensitivity is the simple confidence interval determined by $\sigma(\tau) / \sqrt{N(\tau)}$, where $\sigma(\tau)$ is computed from $N(\tau)$ points.\cite{Riley08} The systematic uncertainty in the sensitivity arises from correlated noise in the sensitivity data, altering the -1/2 white noise scaling of the ADEV at $\tau = 1$ second. After fitting a -1/2 slope to the initial 20-100 points of the ADEV curve (\cref{fig:GMOT_LPAI_Allan}b), we estimate this systematic uncertainty as the absolute deviation between the fit at $\tau = 1$ second and the computed $\sigma(1)$. The error bars in \cref{fig:GMOT_LPAI_Allan} and \cref{fig:GMOT_LPAI_Perf} represent the quadrature sum of statistical and systematic uncertainties. 

With the above-baseline sensitivity and bias stability values in hand, we next explored how the LPAI performance depends on data rate and interrogation time. For data rate from 20 to \SI{100}{Hz}, we measured acceleration sensitivities using ADEV analysis. \cref{fig:GMOT_LPAI_Perf}a shows the measured sensitivity as a function of data rate. At each data rate, we then optimized $T$ (\cref{fig:GMOT_LPAI_Perf}b). The results reveal a trade-off: high data rate improves sensitivity via statistical averaging ($S \propto 1/\sqrt{R}$) for a given time, while low data rate enhances per-shot sensitivity by enabling longer interrogation times ($S \propto 1/T^2$). At data rate below the optimum, fewer measurement cycles reduce the averaging, which outweighs the benefit of a longer $T$, degrading overall sensitivity; in this regime the LPAI's lower low-frequency cutoff ($\simeq \frac{1}{2T}$) also makes it more susceptible to vibration noise, while its high-frequency cutoff is set by the $\pi/2$ pulse duration $\tau_{\pi/2}$. Conversely, above the optimum data rate, sensitivity is dominated by increased sampling, and the $1/T^2$ scaling only plays a secondary role. 

By employing circularly polarized Raman beams in a retro-reflection geometry (see Experimental Setup and Methods), we were able to achieve the maximal data rate as \SI{100}{Hz}. Using our compact GMOT-based LPAI, we achieved sensitivities ranging from \SI{550}{\nano g / \sqrt{Hz}} (\SI{20}{Hz}) to \SI{1.49}{\micro g / \sqrt{Hz}} (\SI{45}{Hz}) (\cref{fig:GMOT_LPAI_Perf}), with a bias stability of \SI{71}{\nano g} (\cref{fig:GMOT_LPAI_Allan}b).

\section*{Discussion and Outlook}

Quantum IMUs are expected to play a crucial role in guidance, navigation, and control for field-deployable applications because of their potential for high sensitivity and improved long-term stability. However, mission-relevant conditions---including high dynamics, broadband vibrations, acoustic noise, thermal cycling, and radiation---place stringent demands on the underlying atom interferometers and their subsystems. To operate beyond quasi-static or mild-vibration conditions and address high-vibration, high-dynamics maneuvers, it is necessary to leverage the complementary strengths of quantum IMUs, which offer high sensitivity and long-term bias stability, and conventional IMUs, which provide high data rates and large dynamic ranges. Progress toward field deployment also requires advances in several key areas, including compact, rugged physics packages, \cite{Lee22, Seo26} compact PIC laser subsystems, \cite{Kodigala24} sustained high-data-rate LPAI operation, and advanced quantum-classical sensor-fusion algorithms. In this work, we highlight two of these areas---compact GMOT physics packages and high-data-rate GMOT-based LPAI operation---which have been implemented simultaneously to address several major technical limitations.

First, we reduced the conventional six-beam MOT architecture to a single-beam GMOT. The compact, short-baseline GMOT architecture improves robustness against vibration and highly dynamic environments by simplifying optical delivery, reducing the number of discrete optical components, and enabling fixed-optics operation, thereby shrinking the physics-package volume to \SI{\approx 2}{L}. In addition, high-data-rate GMOT-based LPAI operation helps mitigate vibration-induced noise and spatial mismatch between the falling atoms and the interrogation beams. Using this promising platform toward field-deployable applications, we demonstrated GMOT-based cold-atom generation with sub-Doppler cooling and GMOT-based LPAI operation at data rates up to \SI{100}{Hz}. Increasing sampling rate improves sensitivity with statistical advantage and raises the low-frequency cutoff of the atom interferometer transfer function, thereby helping to reject environmental vibration. 

Second, by optimizing the interrogation time at each data rate, we demonstrated sensitivities of \SI{550}{\nano  g / \sqrt{Hz}} at \SI{20}{Hz}, \SI{1.49}{\micro g/\sqrt{Hz}} at \SI{45}{Hz}, and \SI{6.31}{\micro g/\sqrt{Hz}} at \SI{70}{Hz}, with a bias stability of \SI{71}{ng}. This high-data-rate GMOT-based LPAI study demonstrates an unprecedented balance between the statistical gain provided by high data rates and the enhanced sensitivity enabled by longer interrogation times through the characteristic $T^2$ scaling of LPAIs.

Third, using the PIC-compatible laser architecture \cite{Lee22} based on \SI{1560}{nm}-to-\SI{780}{nm} frequency-doubling, we incorporated both a standalone 795-nm optical-pumping laser to increase the atomic population in the desired initial state and a tapered amplifier to boost the GMOT cooling-beam power. These upgrades increase the GMOT atom number and improve recapture efficiency, and they support high-data-rate GMOT-based cold-atom generation with sub-Doppler cooling at rates up to \SI{100}{Hz}. Efficient light delivery to the in-vacuum, microfabricated grating chip is critical: while a large-diameter Gaussian beam truncated by the optical window improves spatial uniformity, it delivers cooling power inefficiently. To increase the delivery efficiency above \SI{70}{\%}, we developed and tested a custom 10-cm-long flat-top beam shaper that expands and collimates the beam, eliminating the need for the tapered amplifier. Looking ahead, we are also developing a heterogeneously integrated PIC-based laser subsystem that incorporate a four-channel silicon-photonic single-sideband modulator chip, consolidating all the functions required for LPAI operation.

Fourth, to overcome limited recapture efficiency imposed by the finite GMOT capture volume, we propose employing helper beams during Doppler cooling (GMOT reload and recapture). In future prototypes, adding helper beams along the horizontal and vertical sensing axes could sustain high recapture efficiency and high sensitivity at data rates above \SI{70}{Hz}. This approach also provides a scalable path toward time-multiplexed operation and multi-axis inertial sensing, supported by distributed detection channels and orthogonal Raman beam geometries. 

Together, our integration of compact physics packages, PIC-compatible/-based laser subsystems, high-data-rate LPAI operation, and forward-looking quantum-classical sensor-fusion strategies represents a clear step toward field-deployable GMOT-based LPAIs that maintain performance under vibration and high dynamic conditions. Such systems could enable practical quantum inertial and gravity sensors for navigation, geophysics, and other real-world applications.

\section*{Experimental Setup and Methods}

In this section, we describe the experimental setup for a GMOT-based LPAI---including the compact GMOT physics package, the laser system, the Raman laser subsystem, and the methods for improving GMOT atom number and atom recapture. 

\subsection*{Compact GMOT Physics Package}

\begin{figure}[t!]
\includegraphics[width=1\columnwidth]{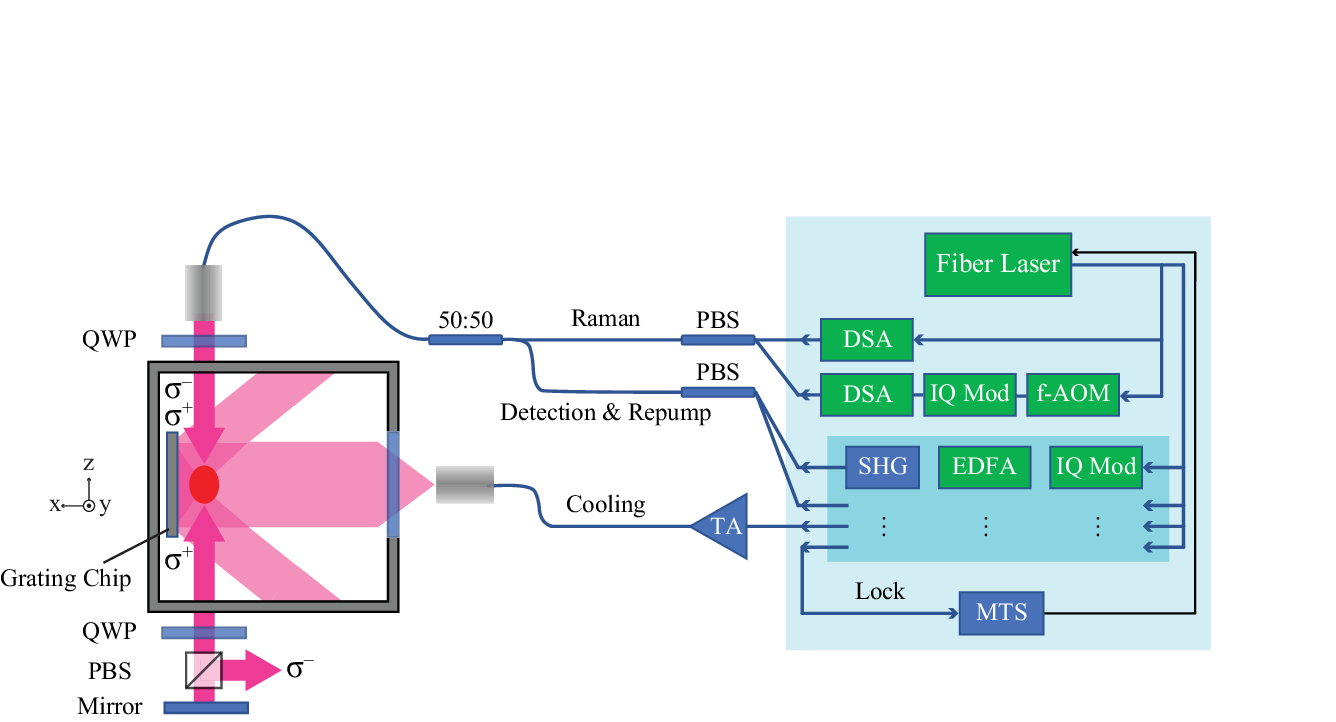}
\caption{Setup diagram of the GMOT physics package and laser systems used in GMOT-based LPAI experiments. Fiber Laser: 1560-nm fiber laser; IQ Mod: single-sideband modulator; SHG: second harmonic generator; MTS: modulation transfer spectroscopy; TA: tapered amplifier; DSA: double-stage amplifier comprising 1560-nm and 780-nm semiconductor optical amplifiers (SOAs) with an interposed frequency-doubling stage; f-AOM: fiber-based acousto-optic modulator.}
\label{fig:Laser_Setup}
\end{figure}

At the heart of the GMOT physics package is a binary microfabricated reflective grating chip mounted in a custom titanium vacuum chamber\cite{Little21} (70 mL, fused silica or quartz window), evacuated to $\sim 5\times10^{-9}$ Torr by a 2 L/s ion pump (\cref{fig:Laser_Setup}). The thin titanium wall and an external mu-metal shield minimize eddy currents and stray fields when the bias-coil currents are switched. Cooling, repump, and detection beams are delivered via polarization-maintaining fiber. For the GMOT cooling beam, we expand the Gaussian beam to a 28-mm $1/e^2$ diameter and use an aperture to truncate it to $\approx19$ mm at the viewport. After passing through a polarizer and quarter waveplate, the beam illuminates the three-segment grating, which diffracts into three first orders at \SI{40.56}{\degree}, \cite{Lee22} forming a robust tetrahedral MOT. All the GMOT optics are actively aligned and permanently fixed in place with epoxy, eliminating moving parts.

In our previous setup, the detection port was fixed and the Raman-beam port held only fixed optics, which hindered precise atom-light alignment. We have now replaced both ports with kinetic mounts, enabling fine alignment of the fluorescence-collection optics and the Raman-beam path for improved performance. Once aligned, the mounts are locked in place for stable data acquisition. 

In our field-deployable design, we added four detection ports to increases the detection volume and support longer interrogation times in LPAI measurements without having to re-optimize port locations for each data rate. Additionally, a custom flat-top beam shaper (10 cm long, $>70 \%$ transmission) can serve as both the GMOT cooling beam and the Raman beams, easing the alignment challenges imposed by fixed optics.  

\subsection*{Laser System}

Our single 1560-nm master laser (NP Photonics FLM-150-3-1560.49-1-S-A; linewidth \SI{\sim 5}{kHz}) provided most of required beams via fiber splitting. We routed its output into five channels (lock, cooling, repump, detection, and Raman); on each channel we performed: (1) IQ-modulation for single-sideband generation, (2) optical amplification by an erbium-doped fiber amplifier (EDFA), and (3) second harmonic generation (SHG) to SI{780}{nm} in a periodically poled lithium niobate (PPLN) waveguide. \cref{fig:Laser_Setup} shows the block diagram. 

The lock channel relied on modulation transfer spectroscopy (MTS) \cite{Noh11} for the $F = 2 \rightarrow F' = 3$ transition on the \Rb D$_2$ line ($5^2S_{1/2} \rightarrow 5^2P_{3/2}$, \SI{780}{nm}) to stabilize the master laser (\cref{fig:Laser_Setup}). By modulating only the pump beam, we suppressed probe-modulation-induced drifts and achieved more robust laser locking than with frequency modulation spectroscopy. \cite{Bjorklund80}

The cooling and repump channels generated MOT beams via an IQ modulator (EO Space IQ-0D6S-25-PFA-PFA-LV), \cite{Zhu18, Rammeloo20} an EDFA (IPG Photonics EAR-2K-C-LP-SF), and an SHG (AdvR, RSH-T0780-P15P78AL0). The detection followed the same architecture. To boost the cooling-beam power, we added a 780-nm tapered optical amplifier; to optimize the fluorescence detection, we remounted the fixed detection optics on a three-axis translation stage so that the fluorescence-collection lens could be positioned optimally for any chosen drop time.

We also employed a $\pi$-polarized $D_1$ optical-pumping beam (quantization axis of \SI{0.665}{G} along the z-axis; see \cref{fig:Laser_Setup}) for the $F=2 \rightarrow F'=2$ transition on the \Rb $D_1$ line ($5^2S_{1/2} \rightarrow 5^2P_{1/2}$, \SI{795}{nm}) to prepare the initial atomic state $|F = 2, m_F = 0\rangle$. The beam propagates along the +y-axis.

\subsection*{Raman Laser Subsystem}
Stimulated Raman transitions in our GMOT-based LPAI require two phase-coherent beams separated by \SI{\sim 6.834}{GHz}. We generated both frequencies in a single fiber and delivered them in a retro-reflection geometry \cite{Gauguet08} to suppress phase noise via a common-mode path (\cref{fig:Laser_Setup}). The single-photon detuning of the Raman beams is \SI{484}{Mz} below the $F=2 \rightarrow F'=0$ transition.

Two cross-linearly-polarized Raman beams are converted into circularly polarized Raman beams using a first quarter waveplate, which suppresses the Doppler-free peak. A second quarter waveplate together with a polarizing beam splitter then removes the unwanted counter-propagating beam pair before the retro-reflection mirror, thereby filtering out the other Doppler-sensitive peak. This direct addressing of the Doppler-sensitive resonance---without detuning it from the Doppler-free transition---doubling our maximum data rate to \SI{100}{Hz}. The Raman beams' quantization axis of \SI{0.665}{G} is along the z-axis (see \cref{fig:Laser_Setup}).

To implement active intensity stabilization for the Raman-field double-stage semiconductor optical amplifier (SOA) system, we applied current feedback to the SOA. To implement Mach-Zehnder matter-wave interference, we employed the light-pulse sequence $\pi/2 \rightarrow T \rightarrow \pi \rightarrow T \rightarrow \pi/2$, which split, redirected, and recombined matterwave packets, imprinting a phase $\Phi \approx k_{\rm eff}\,g\,T^2$. We scanned the phase $\varphi$ of the final $\pi/2$ pulse from 0 to $2\pi$ (\cref{fig:GMOT_LPAI_Allan}a) to locate the mid-fringe, then locked the phase there and recorded interferometer data for Allan-variance analysis (\cref{fig:GMOT_LPAI_Allan}b).

\subsection*{Methods for Improving GMOT Atom Number and Atom Recapture}

High-data-rate GMOT operation requires precise timing and amplitude control of the cooling, repump, optical pumping, and detection beams. The cooling ($F = 2 \rightarrow F' = 3$, \SI{780}{nm}) and repump ($F = 1 \rightarrow F' = 2$, \SI{780}{nm}) beams load atoms into the GMOT and perform sub-Doppler cooling; the $\pi$-polarized $D_1$ optical-pumping beam ($F = 2 \rightarrow F' = 2$, \SI{795}{nm}) prepares the initial state; and fluorescence detection (with repump) yields normalized atom number signals. The detection beam is on resonance with the $F = 2 \rightarrow F' = 3$ transition (\SI{780}{nm}). The beam diameter of the repump, detection, and Raman beams is \SI{5.42}{mm} ($1/e^2$ Gaussian diameter).

For normalized detection, an initial 100-\SI{}{\micro s} pulse measures the $F = 2$ population ($T_\text{F=2}$ = \SI{100}{\micro s}). A second 100-\SI{}{\micro s} pulse, with repump, transfers all atoms into the $F = 2$ and measures the total population ($T_\text{F=1+2}$ = \SI{100}{\micro s}). Peak steady-state atom number occurred at $I_\text{cool} = 7.99 \,\rm{mW/cm^2}$ and a detuning of $-3.33\,\Gamma$. 

The detection peak intensity is $I_\text{det} =$ \SI{1.1}{mW/cm^2}.
The repump peak intensity is $I_\text{repump} =$ \SI{41.6}{mW/cm^2}.
The cooling peak intensity is $I_\text{cool} =$ \SI{10.6}{mW/cm^2}.

To boost the GMOT atom number, we seeded a 780-nm tapered amplifier (MOGLabs MOAL) with 30 mW and achieved to 3 W free-space output (1.5 W fiber-coupled). We calibrated our avalanche photodiode-based fluorescence detection against a charge-coupled device camera and measured $\sim2.2\times10^6$ atoms at equilibrium.

\begin{table}[h!]
\centering
\begin{center}
\begin{tabular}{l l c}
  \hline
  Sequence & Description & Duration  \\ 
  \hline
                      & Experiment sync pulse & \SI{5}{\micro s} \\ 
  \hline
   Loading MOT        & Set bias fields for loading   & \SI{0}{\micro s} \\
                      & Wait for cooling IQ to settle & \SI{200}{\micro s} \\
                      & Time for loading MOT          & 0 to \SI{2}{s} \\
  \hline
  Sub-Doppler cooling & Zero bias fields    & \SI{700}{\micro s} \\
                                            & Sub-Doppler cooling  & 2 to \SI{2.5}{ms} \\
  \hline
  State Preparation   & Set quantization axis                  & \SI{500}{\micro s} \\ 
                      & $D_1$ optical pumping ($T_\text{op}$)  & \SI{10}{\micro s} \\
  \hline
  Light-Pulse Sequence & $\frac{\pi}{2}$ pulse  & \SI{1.1}{\micro s} \\ 
                      & Interrogation time T    & 0.1 to \SI{10}{ms} \\ 
                      & $\pi$ pulse             & \SI{2.2}{\micro s} \\ 
                      & Interrogation time  T   & 0.1 to \SI{10}{ms} \\ 
                      & $\frac{\pi}{2}$ pulse   & \SI{1.1}{\micro s} \\
  \hline
  Detection & Wait for detection IQ to settle & \SI{200}{\micro s} \\
                       & Software specific delay  & \SI{100}{\micro s} \\
                       & Detection pulse          & \SI{100}{\micro s} \\
                       & Software specific delay  & \SI{100}{\micro s} \\
                       & Detection pulse          & \SI{100}{\micro s} \\
  \hline
                       & Wait time before next cycle & \SI{100}{\micro s} \\ 
  \hline
\end{tabular}
\caption{Sequence of a high-data-rate, high-sensitivity GMOT-based LPAI. The range of times for loading GMOT and delay differ depending on the data rate.}
\label{tab:GMOT_LPAI_seq}
\end{center}
\end{table}


\begin{acknowledgements}
We would like to express our gratitude to Cody Bassett, Roger Ding, Aaron Ison, Kyle Fuerschbach, Elliot Diamond-Hitchcock, Christian Sanchez, Matthew Huh, Grant Biedermann, and Peter Schwindt for their support and helpful discussions. This work was supported by the Naval Air Warfare Center Aircraft Division Lakehurst (NAWCAD LKE) under Contract No. N68335-24-C-0546 and by the Laboratory Directed Research and Development program at Sandia National Laboratories. Sandia National Laboratories is a multimission laboratory managed and operated by National Technology and Engineering Solutions of Sandia, LLC, a wholly owned subsidiary of Honeywell International Inc., for the U.S. Department of Energy’s National Nuclear Security Administration under contract DE-NA0003525. Any opinions, findings, and conclusions, or recommendations expressed in this material are those of the author(s) and do not necessarily reflect the views of the NAWCAD LKE, the U.S. Department of Energy, or the United States Government. This paper describes objective technical results and analysis; any subjective views or opinions expressed herein do not necessarily represent those of the funding agencies.
\end{acknowledgements}

\section*{Conflict of Interest}
The authors declare no conflicts of interest.

\section*{Data Availability}
Data underlying the results presented in this paper are not publicly available at this time but may be obtained from the authors upon reasonable request.
 
\nocite{*}

\bigskip


\end{document}